\documentclass[final]{pasj01}

\usepackage{graphicx}

\begin{document} 
\Received{2018/05/28}
\Accepted{2018/07/31}

\title{On Surface Brightness and Flux Calibration for Point and Compact Extended Sources in the AKARI Far-IR All-Sky Survey (AFASS) Maps}

\author{Toshiya \textsc{Ueta}\altaffilmark{1}}%
\altaffiltext{1}{Department of Physics and Astronomy, University of Denver, Denver, CO 80208, U.S.A.}
\email{toshiya.ueta@du.edu}

\author{Ryszard \textsc{Szczerba}\altaffilmark{2}}
\altaffiltext{2}{Nicolaus Copernicus Astronomical Centre, PAS, ul.\ Rabia\'{n}ska 8, 87-100 Toru\'{n}, Poland}

\author{Andrew G.\ \textsc{Fullard}\altaffilmark{1}}

\author{Satoshi \textsc{Takita}\altaffilmark{3,4}}
\altaffiltext{3}{Institute of Space and Astronautical Science, Japan Aerospace Exploration Agency, 3-1-1 Yoshinodai, Sagamihara, Kanagawa 252-5210, Japan}
\altaffiltext{4}{Presently: National Astronomical Observatory of Japan, 2-21-1 Osawa, Mitaka, Tokyo 181-8588, Japan}


\maketitle

\begin{abstract}
The {\sl AKARI\/} Infrared Astronomical Satellite produced the all-sky survey (AFASS) maps in the far-IR at roughly arc-minute spatial resolution, enabling us to investigate the whole sky in the far-IR for objects having surface brightnesses greater than a few to a couple of dozen MJy\,sr$^{-1}$.
While the AFASS maps are absolutely calibrated against large-scale diffuse emission, it was uncertain whether or not an additional flux correction for point sources was necessary.
Here, we verify that calibration for point-source photometry in the AFASS maps is proper.
With the aperture correction method based on the empirical point-spread-function templates derived directly from the AFASS maps, fluxes in the {\sl AKARI\/} bright source catalogue (BSC) are reproduced.
The {\sl AKARI\/} BSC fluxes are also satisfactorily recovered
with the 1-$\sigma$ aperture, which is the empirical equivalent of an infinite aperture.
These results confirm that in the AFASS maps far-IR photometry can be properly performed by using the aperture correction method for point sources and by summing all pixel values within an appropriately defined aperture of the intended target (i.e., the aperture photometry method) for extended sources.
\end{abstract}

\section{Introduction}

The {\sl AKARI\/} Infrared Astronomical Satellite ({\sl AKARI\/}; \cite{akari}) is the Japanese infrared space mission launched on 2006 February 21 (UT).
The primary mission of {\sl AKARI\/} was to conduct an all-sky survey in the mid- and far-infrared at higher spatial resolutions and over a wider spectral range than its predecessor, the Infrared Astronomy Satellite ({\sl IRAS}; \cite{Neugebauer_1984}). 
For this purpose, {\sl AKARI\/} was outfitted with a cryogenically-cooled telescope with a diameter of 68.5\,cm and two instruments, the Far-Infrared Surveyor (FIS; \cite{Kawada_2007}) and the Infrared Camera (IRC; \cite{Onaka_2007}), covering a wavelength range of 2 -- 180\,$\micron$.
{\sl AKARI\/} carried out its 550-day cryogen mission until it exhausted liquid Helium on 2007 August 26, and continued its post-cryogen mission in the near-infrared until the satellite was finally switched off on 2011 November 24.

The FIS instrument covers the wavelength range of 50--180$\,\mu$m with two sets of Ge:Ga arrays, the Short-Wavelength (SW) and Long-Wavelength (LW) detectors in the N60 (50 -- 80\,$\micron$) and WIDE-S (60 -- 110$ \mu$m) bands and the WIDE-L (110 -- 180$\,\mu$m) and N160 (140 -- 180$ \mu$m) bands, respectively \citep{Doi_2002,Fujiwara_2003}. 
Pixel scales in the detectors were designed to be similar to the diffraction limit of the telescope, at around $30\arcsec$.
For the all-sky survey, the natural 100-minute sun-synchronous orbit of {\sl AKARI\/} was used to achieve a scan speed of 3\farcm6\,s$^{-1}$ during the survey period from 2006 April to 2007 August, allowing to cover over 96 per cent of the sky with two or more scans, and in fact resulted in 99 per cent sky coverage \citep{Doi_2015,Takita_2015}.

The primary absolute surface-brightness calibration of the FIS instrument was done through 
(1) pre-launch laboratory measurements of a blackbody source which indicated a 5\,$\%$ accuracy, and
(2) on-orbit measurements of infrared cirrus regions without significant small-scale structures and comparisons of the results between FIS and the DIRBE instrument on-board the {\sl COBE} satellite \citep{Matsuura_2011}. 
The additional absolute calibration of the {\sl AKARI} far-infrared all-sky survey (AFASS) images was done, especially for very bright regions along the Galactic plane, via iterative comparison between the surface brightnesses in the AFASS images and expected surface brightnesses based on the DIRBE zodi-subtracted mission average (ZSMA) data set \citep{Hauser_1998} augmented with the Gorjian zodiacal emission model \citep{Gorjian_2000}.
The resulting uncertainties were determined to be 
less than 10\,\% for surface brightnesses greater than $10, 3, 25$, and 26\,MJy\,sr$^{-1}$ at the N60, WIDE-S, WIDE-L,and N160 bands, respectively, at the spatial resolution of DIRBE at approximately half a degree \citep{Boggess_1992,Takita_2015}.

Hence, the presently archived AFASS mapping data (the Public Release Version 1, AFASSv1 hereafter)\footnote{In the Data ARchives and Transmission System (DARTS) maintained by ISAS/JAXA (http://www.darts.isas.jaxa.jp/astro/akari/).} should give correct surface brightnesses of diffuse background emission.
However, when aperture photometry was performed for a set of infrared flux calibrators detected in the AFASS maps in the N60, WIDE-S, and WIDE-L bands, the resulting fluxes came out to be roughly 30 -- 60\,\% less than expected \citep{Arimatsu_2014}.
In addition, \citet{Takita_2015} compared fluxes listed in the {\sl AKARI\/}/FIS All-Sky Bright Source Catalogue Ver.\,1 (BSCv1; \cite{Yamamura_2009}) and those measured from the corresponding sources in the AFASS images via aperture photometry (with the 90$\arcsec$ radius aperture with the 120--300$\arcsec$ radius sky annulus) and noted that the ratio of the AFASS-to-BSC fluxes as a function of the number of sources within 5$\arcmin$ of the detected source (NDENS) was found to become large for $\mathrm{NDENS} \ge 1$ because of contamination by nearby sources.
Note that the absolute flux calibration for the BSCv1 fluxes was done directly from the time-series detector signal readouts of photometry reference objects without making surface brightness maps \citep{Yamamura_2009}.

As for the origin of the observed discrepancy between the observed and expected fluxes of point sources, \citet{Arimatsu_2014} offered two possible explanations:
(1) the observed point source fluxes are always underestimated because of the extended point-spread-function (PSF) component beyond the aperture radius (e.g., \cite{Arimatsu_2011}), and
(2) the sensitivity of the FIS detector may be lower for point sources than for diffuse sources because the signal from the point sources can be missed due to the slow transient of the FIS detector \citep{Shirahata_2009,Ueta_2017}. 
However, 
\citet{Doi_2015} discussed in detail about the slow transient effects in the AFASS data and how they were mitigated and removed.
Thus, 
the observed flux underestimates is more likely caused by the unobserved extended point-spread-function (PSF) component beyond the 90$\arcsec$ aperture radius.

Thus, in the present work, 
we examine if the missing extended surface brightness component is truly the cause of the observed flux underestimates for point sources.
\citet{Arimatsu_2014} and \citet{Takita_2015} previously measured fluxes of stars in the Cohen catalogue \citep{Cohen_1999}, which contains IR photometric standard stars with a flux range of 0.02 --10\,Jy in the WIDE-S band.
Because of the relative faintness of these Cohen standard stars, many of them fell outside of the {\sl AKARI\/} detection limit and their subsequent analyses were conducted using a stacking method to improve on the signal-to-noise ratio (S/N).
Here, we take a different approach
to verify the point-source flux calibration of the AFASS images.
First, we establish the empirical PSF of the AFASS images in all four bands from the AFASS maps themselves (\S\,\ref{S:psf}).
Then, we perform point-source photometry using the aperture correction factors derived from the empirical PSF templates
as well as contour photometry with the 1-$\sigma$ aperture that simulates an infinite aperture (\S\,\ref{S:photom}),
before summarizing the entire work (\S\,\ref{S:summary}).

\section{Establishing the Empirical PSFs in the AFASS Images}
\label{S:psf}

\subsection{Source Selection}

First, we need to identify a set of point sources in the AFASS images as PSF/photometric references to perform photometry and assess the quality of flux calibration for point sources.
\citet{Arimatsu_2014} and \citet{Takita_2015} adopted a set of 352 photometric standard stars \citep{Cohen_1999} for their analysis.
However, because these Cohen stars are usually used as optical and near-IR photometry standard stars,
they are not intrinsically bright in the far-IR (their expected flux is less than 10\,Jy in the WIDE-S band).
Because \citet{Arimatsu_2014} and \citet{Takita_2015} adopted only those that were detected at 2.5\,$\sigma$ or greater in their analysis, they were able to use only 49, 97 and 9 stars in the N60, WIDE-S and WIDE-L bands, respectively, and unable to find any suitable stars in the N160 band.
Also, typical far-IR photometric standards, such as asteroids and planets \citep{Muller_2002}, are not available in the all-sky survey scan AFASS images (i.e., removed as moving objects). 
Hence, we need to take a different approach to increase the number of sources that can serve as far-IR photometry standard stars in our analysis.

Our aim here is to provide good-quality point sources that are reasonably bright.
Thus, rather than starting from some existing list of sources, we opt to create our own list of sources for which source detection is guaranteed in the AFASS images.
To do so, we select sources from the {\sl AKARI} FIS Bright Source Catalog Ver.\,2 (BSCv2; \cite{Yamamura_2016})\footnote{Available via http://www.ir.isas.jaxa.jp/AKARI/Archive/.} using the following two criteria, 
(1) ${\rm FQUAL} = 3$ (the presence of the source is confirmed and its flux determined to be valid) in all four FIS bands and 
(2) ${\rm NDENS} \le 2$ (there are at most two nearby objects within $5\arcmin$ of the source).
The BSCv2 release notes state that ${\rm FQUAL}=3$ and ${\rm GRADE}=3$ is the basic condition to select good quality sources
\citep{Yamamura_2016}.
${\rm GRADE}=3$ means that 
${\rm FQUAL}=3$ in two or more bands, or
${\rm FQUAL}=3$ in one band and ${\rm NSCANC} \ge 4$ (the number of scans in which the source is detected).
The first criterion that we adopt is therefore more stringent than the release notes recommend.
This initial filtering identifies 658 well-detected relatively isolated sources.

\begin{figure}
\begin{center}
    \includegraphics[width=\columnwidth]{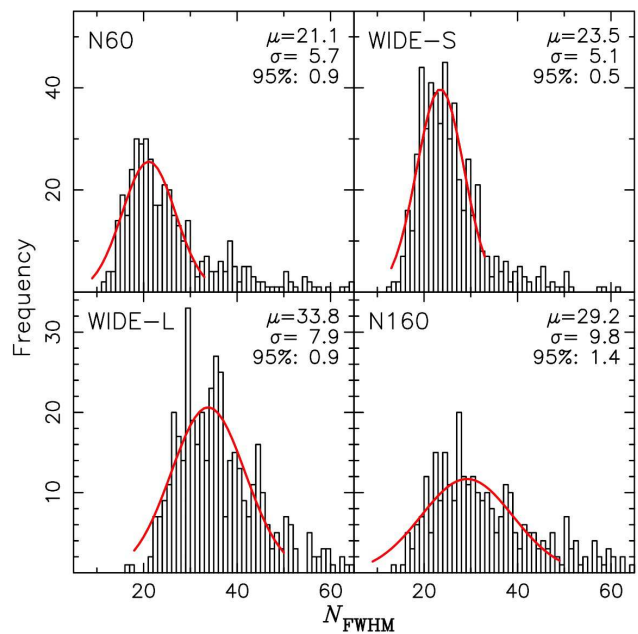}
\end{center}
\caption{%
The distribution of the FWHM pixel counts for the 658 BSCv2 sources in each of the four FIS bands as indicated at the top-left of the frame.
The Gaussian fit for each distribution is drawn by a gray curve, and the mean ($\mu$), standard deviation ($\sigma$), and 95\,\% confidence half-interval (95\,\%) are shown at the top-right of the frame.}\label{F:peakpix}
\end{figure}

Then, we cut out $20\arcmin \times 20\arcmin$ images centered at the coordinates of the 658 BSCv2 sources from the archived AFASS images\footnote{Each of the archived AFASS images covers a $6\fdg0 \times 6\fdg0$ region of sky in the ecliptic coordinates spaced by an increment of $5\fdg0$.}
only when
(1) an emission peak exists at the reported BSC coordinates with greater than 20 and 10\,$\sigma$ detection in the wide (WIDE-S and WIDE-L) and narrow (N60 and N160) bands, respectively,
(2) the emission peak of the BSCv2 source can be fit with a 2-D Gaussian
(as the AFASS maps are constructed assuming a Gaussian beam profile; \cite{Doi_2015}), and
(3) no brighter sources exist in the same $20\arcmin \times 20\arcmin$ cutout region.

Because these are BSCv2 sources, they are expected to be point sources (or point-like, at least).
However, there is no guarantee that they are actually point sources.
Hence, we count the number of pixels within the full-width at half-maximum (FWHM) 
of the peak, and determine the distribution of the FWHM pixel counts for these 658 sources in each of the four FIS bands (Fig.\,\ref{F:peakpix}).
The Gaussian fit to the distribution reveals the mean FWHM size (in pixel counts), standard deviation, and 95\,\% confidence interval.
We consider those that fall within the 95\,\% confidence interval of the mean FWHM size (73, 78, 64, and 45 sources in the N60, WIDE-S, WIDE-L, and N160 bands, respectively) as representative of point sources.
The nature of these point sources is almost exclusively distant galaxies (about 88\,\%), and the rest is split between stellar and unidentified objects with about equal proportions, according to their designations in the SIMBAD database\footnote{http://simbad.u-strasbg.fr/simbad/} (cross-matched by the coordinates).

\begin{figure*}
\includegraphics[width=\textwidth]{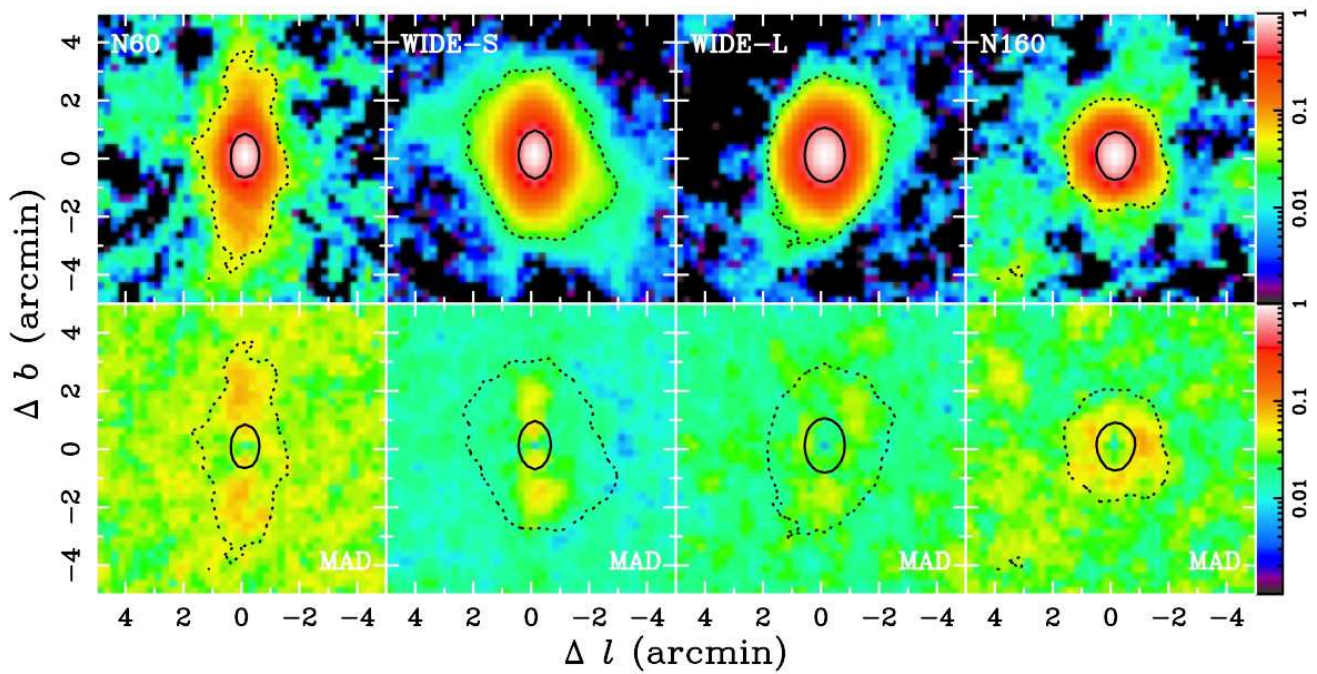}
\caption{\label{F:psfmad}%
The {\sl AKARI\/}/FIS super-PSF images (top row) and  MAD maps (bottom row) in the N60, WIDE-S, WIDE-L and N160 bands (top row; from left to right).
The logarithmic color scaling of the images, from 0.1\,\% to 100\,\% relative to the peak intensity, is indicated in the wedge on the right.
The solid and dotted contours represent the FWHM and 5\,$\sigma$ detection level, respectively.
The image orientation follows the ecliptic coordinates, which are adopted for the original AFASS maps.}
\end{figure*}

\subsection{Super-PSF Images\label{psf}}

Using the AFASS cutout images of the BSCv2 point sources selected as above, we compute the normalised median ``super-PSF" image and the corresponding median absolute deviation (MAD) map for each of the four {\sl AKARI\/} bands (Fig.\,\ref{F:psfmad}).
When compiling these super-PSF images, the emission peak is fit with a 2-D Gaussian function in each of the selected PSF reference images so that the PSF images are properly aligned (i.e., shifted and de-rotated) before being normalised and co-added.
The MAD image is made by taking the median of the absolute difference between each PSF image and the super-PSF image.
The MAD maps, therefore, graphically indicate how individual PSF images are statistically different (or the same) from the corresponding super-PSF image at each pixel.

\begin{table}
\tbl{\label{T:psf}%
Super-PSF FWHM size of the {\sl AKARI\/}/FIS all-sky survey map} {%
 \begin{tabular}{lccc}
  \hline
  Band & in-scan FWHM & cross-scan FWHM & Eccentricity \\
  & (arcsec) & (arcsec) & \\
  \hline
  N60         & $\phantom{1}96.4\pm7.9$ & $56.7\pm4.1$ & 0.81\\
  WIDE-S  & $103.8\pm5.1$ & $68.5\pm3.1$ & 0.75\\
  WIDE-L   & $104.2\pm7.9$ & $77.9\pm7.0$ & 0.66\\
  N160       & $\phantom{1}85.5\pm8.3$ & $73.3\pm5.8$ & 0.43\\
  \hline
 \end{tabular}}
\end{table}

Within the region that registers more than 5-$\sigma$ detection (the dotted contour in Fig.\,\ref{F:psfmad}), the median MADs intrinsic to the source emission are determined to be 
$4.4\pm1.4\,\%$,
$1.5\pm0.6\,\%$,
$2.4\pm0.6\,\%$, and
$4.2\pm1.4\,\%$,
for the N60, WIDE-S, WIDE-L and N160 bands, respectively.
These values indicate that the shape of the adopted point sources is identical more than 95\,\% in each of the {\sl AKARI\/} bands, i.e., the PSF shape is fairly uniform.
This uniformity among the adopted BSCv2 point sources implies that the shape of the PSF does not vary with the source brightness (i.e., the PSF is scale invariant; e.g., \cite{Ueta_2017}), providing the rationale to establish the super-PSF images by stacking point sources of various fluxes. 
In addition, comparisons with the Cohen star PSFs  \citep{Arimatsu_2014} show less than 2\,\% differences on average between the surface brightness distributions in the PSF core (i.e., within FWHM).
Therefore, we adopt these super-PSF images as representative of the PSF structure in the AFASS maps.

The super-PSF shape appears more elongated along the in-scan direction (i.e., along the ecliptic latitude) than along the cross-scan direction (i.e., along the ecliptic longitude) and the degree of elongation is seen more strongly in the bands at shorter wavelengths.
The elongation is exhibited in both toward the ecliptic N and S directions, because sometimes the object is scanned from N to S and other times from S to N.  
The measured PSF/FWHM elongations and eccentricity are listed in Table\,\ref{T:psf}. 
Overall, the PSF structure in the AFASS maps based on images of the BSCv2 point sources is consistent with that based on the Cohen standard stars (Table\,2 of \cite{Takita_2015}).
The FWHM sizes of the super-PSFs are found to be larger than those found by \citet{Takita_2015} for the stacked Cohen stars (1.5\,\% to 81\,\% with the overall average 24.7\,\%, 11.4\,\% and 37.9\,\% for in-scan and cross-scan FWHMs, respectively).
This is because the point-source FWHM size for the present analysis is determined by sheer statistics, i.e., by the most populous point-like sources (Fig.\,\ref{F:peakpix}), which are not necessarily genuine point sources by nature.
Also, because the present analysis deals with much brighter sources than the Cohen standards used in the previous analyses (up to an order of magnitude brighter), these stars would appear intrinsically larger.

\section{Flux Correction for Point and Compact Extended Sources in the AFASS Maps}
\label{S:photom}

\subsection{Photometry of the Adopted Point Sources}

In their previous analyses, \citet{Arimatsu_2014} and \citet{Takita_2015} used an aperture of 90$\arcsec$ radius and a sky annulus of 180$\arcsec$ width (defined between 120$\arcsec$ and 300$\arcsec$ from the target coordinates) to perform photometry with Cohen stars of their choosing.
As the uncorrected-AFASS-to-BSC flux ratio, they found a constant-value of $0.63\pm0.03$, $0.70\pm0.01$ and $0.38\pm0.04$ in the N60, WIDE-S and WIDE-L bands, respectively.
These factors were implicitly regarded as divisible photometry correction factors when photometry is done with a fixed 90$\arcsec$-radius aperture.
For the N160 band, the Cohen stars turned out to be too dim to allow any assessment.

Because we have now established the PSF shape in each of the four FIS bands of the AFASS, we can perform aperture photometry corrected for by the aperture correction factor based on the measured PSF shape.
Using the super-PSF images and their 2-D Gaussian fits, we calculate the divisible aperture correction factors as a function of the aperture size (Table\,\ref{T:encircle}).
These factors are determined based on the amount of encircled energy measured within a specific aperture defined by the relative surface brightness with respect to the peak intensity.
Instead of a fixed aperture of 90$\arcsec$ radius, 
we employ the FWHM aperture (i.e., defined to be at the $50\,\%$ surface brightness relative to the peak) to measure the flux as the sum of pixel values within the FWHM aperture.
Measured fluxes are then corrected for by the appropriate divisible correction factors (the 50\,\% row in Table\,\ref{T:encircle}) to recover the total flux for an infinite aperture.

\begin{table}
\scriptsize
\tbl{\label{T:encircle}%
Aperture correction factors}{%
\begin{tabular}{rcccc}
  \hline
  Aperture$^{*}$ & & & & \\
  (\%) & N60 & WIDE-S & WIDE-L & N160 \\
  \hline
90 & $ 0.071  \pm 0.002  $ & $ 0.093  \pm 0.001  $ & $ 0.084  \pm 0.001  $ & $ 0.088  \pm 0.001$  \\
80 & $ 0.208  \pm 0.003  $ & $ 0.187  \pm 0.003  $ & $ 0.176  \pm 0.001  $ & $ 0.186  \pm 0.002$  \\
70 & $ 0.258  \pm 0.004  $ & $ 0.240  \pm 0.003  $ & $ 0.301  \pm 0.002  $ & $ 0.318  \pm 0.003$  \\
60 & $ 0.311  \pm 0.005  $ & $ 0.366  \pm 0.004  $ & $ 0.406  \pm 0.002  $ & $ 0.386  \pm 0.004$  \\
50 & $ 0.426  \pm 0.006  $ & $ 0.419  \pm 0.005  $ & $ 0.463  \pm 0.003  $ & $ 0.493  \pm 0.005$  \\
40 & $ 0.497  \pm 0.007  $ & $ 0.510  \pm 0.005  $ & $ 0.559  \pm 0.003  $ & $ 0.576  \pm 0.007$  \\
30 & $ 0.546  \pm 0.008  $ & $ 0.597  \pm 0.006  $ & $ 0.646  \pm 0.004  $ & $ 0.668  \pm 0.008$  \\
20 & $ 0.652  \pm 0.009  $ & $ 0.682  \pm 0.007  $ & $ 0.751  \pm 0.005  $ & $ 0.769  \pm 0.010$  \\
10 & $ 0.737  \pm 0.012  $ & $ 0.757  \pm 0.008  $ & $ 0.853  \pm 0.006  $ & $ 0.871  \pm 0.012$  \\
8 & $ 0.766  \pm 0.013  $ & $ 0.792  \pm 0.008  $ & $ 0.861  \pm 0.006  $ & $ 0.893  \pm 0.013$  \\
5 & $ 0.800  \pm 0.013  $ & $ 0.830  \pm 0.008  $ & $ 0.911  \pm 0.006  $ & $ 0.921  \pm 0.014$  \\
1 & $ 0.895  \pm 0.017  $ & $ 0.913  \pm 0.009  $ & $ 0.978  \pm 0.008  $ & $ 0.986  \pm 0.015$  \\
  \hline
\end{tabular}}
\begin{tabnote}
 $^{*}$The aperture size is defined by the percentage of the peak intensity.
 \end{tabnote}
\end{table}

The aperture-corrected fluxes of the point sources measured in our analysis are then compared against the expected fluxes listed in BSCv2.
This comparison, therefore, would verify the point-source flux calibration of AFASSv1 relative to that of BSCv2.
In other words, our assessment checks for internal consistency between two {\sl AKARI\/} data products.
With very limited access to far-IR photometry standard objects such as planets and asteroids in the public AFASS maps, 
the absolute calibration is beyond the scope of the present study and 
we opt to seek internal consistency between AFASSv1 and BSCv2 in the present analysis. 

In addition to the above aperture-corrected photometry using the FWHM aperture, we perform uncorrected photometry using the 1\,$\sigma$ aperture (i.e., defined to be at the surface brightness equal to the sky $\sigma$) for each of the adopted point-source images at each band for comparison, because this is practically equivalent to doing photometry with the idealised infinite aperture for extended sources.
Furthermore, for the sake of comparison, we perform photometry using the same 90$\arcsec$ aperture with the 120--300$\arcsec$ sky annulus as was done by \citet{Arimatsu_2014} and \citet{Takita_2015}.

\subsection{Assessing the Need for Flux Correction for Non-Diffuse Sources}

\subsubsection{Photometry with a Fixed 90$\arcsec$-Radius Aperture}

First, we compare our photometry results using a fixed 90$\arcsec$-radius aperture against the previous results \citep{Arimatsu_2014,Takita_2015}.
The median of the uncorrected-AFASS-to-BSCv2 flux ratios in each band is found to be 
$0.67 \pm	0.08$,
$0.70 \pm	0.07$,
$0.68 \pm	0.11$, and
$0.69 \pm	0.19$,
for the N60, WIDE-S, WIDE-L, and N160 bands, respectively.
These values indicate that roughly 70\,\% of the expected flux are accounted for within the fixed 90$\arcsec$-radius aperture in all bands.
We see excellent agreement in the SW bands with respect to the previous results ($0.63\pm0.03$ and $0.70\pm0.01$ in the N60 and WIDE-S bands, respectively), 
while our measurements are much better than previously reported in the WIDE-L band ($0.38\pm0.04$).
\citet{Takita_2015} reported an increasing trending in the uncorrected-AFASS-to-BSCv2 flux ratio with the increasing BSCv2 flux in the N60 band for objects less than 1\,Jy, which was not seen in their preliminary analysis \citep{Arimatsu_2014} due to their use of the preliminary data.
In the WIDE-L band, the BSCv2 sources used in our analysis are brighter (5.2 -- 27.5\,Jy) than Cohen stars adopted previously (less than 3\,Jy).
Thus, there may be a similar increasing trending in the WIDE-L band beyond 3\,Jy which was not recognized previously by \citet{Takita_2015}.

In fact, a fixed 90$\arcsec$ radius aperture of the super-PSFs would translate to apertures defined by the surface brightness at 
$20 \pm 9$\,\%,
$27 \pm 8$\,\%,
$34 \pm 8$\,\%, and
$28 \pm 7$\,\%
of the peak intensity 
at N60, WIDE-S, WIDE-L, and N160, respectively.
Each of these would further translate to the aperture correction factor of 0.7 to 0.57, i.e., the 90$\arcsec$-radius apertures would typically cover 60 to 65\,\% of the total flux across the bands (Table\,\ref{T:encircle}).
This is consistent with the median uncorrected-AFASS-to-BSCv2 flux ratio being about 68\,\% across the bands. 
Hence, we consider that our results are robust and confirm that the previously reported flux underestimates at around 40\,\% was caused by the use of the fiducial 90$\arcsec$ aperture that did not account for the surface brightness component beyond the aperture.

\subsubsection{Photometry with the Derived Aperture Correction}

\begin{figure}
\includegraphics[width=\columnwidth]{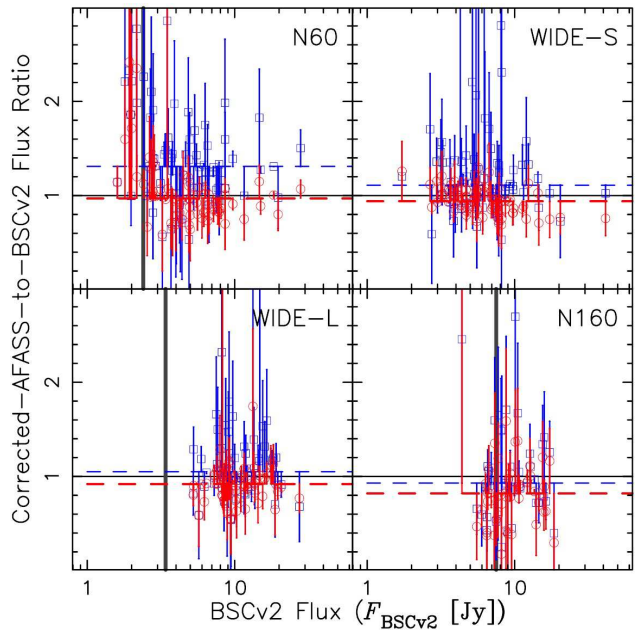}
\caption{\label{F:corrphot}%
The corrected-AFASS-to-BSCv2 flux ratio vs.\ the BSCv2 flux for each of the four {\sl AKARI\/}/FIS bands as indicated at the top-right corner.
Red circles show the ratios based on the aperture correction method, 
while blue squares are those based on the 1-$\sigma$ aperture method.
Corresponding error bars in the ratio are also shown.
The black solid horizontal line shows the unity ratio, whereas 
the red and blue dashed lines indicate the median ratio for the aperture correction and 1-$\sigma$ aperture methods, respectively.
The relatively large scatters in the flux ratio are also previously reported by \citet{Takita_2015} and was attributed to the difference in the actual photometry methods with the AFASS maps and in the BSC.
The vertical thick gray lines indicate the BSCv2 detection limit \citep{Yamamura_2016} at 2.4, 0.44, 3.4, and 7.5\,Jy in the N60, WIDE-S, WIDE-L, and N160 bands, respectively (where visible).}
\end{figure}

Second, we compare our photometry results using 
the FWHM (50\,\% peak) aperture corrected for by the corresponding correction factor.
With the aperture correction method, the corrected-AFASS-to-BSCv2 flux ratio is found to be 
$0.97 \pm	0.14$,
$0.94 \pm	0.10$,
$0.92 \pm	0.10$, and
$0.82 \pm	0.22$.
The derived ratio as a function of the BSCv2 flux for each band is shown in Fig.\,\ref{F:corrphot}.

Here we confirm that the BSCv2 fluxes for the adopted point sources are reproduced within the margin of error by means of the aperture correction method.
As we saw in the previous section, the use of a 90$\arcsec$-radius aperture would generally result in about 40\,\% flux underestimates while this 90$\arcsec$-radius aperture would require flux correction by a divisible factor of about 0.6 (Table\,\ref{T:encircle}).
Raw fluxes are underestimated because of an extended PSF component beyond the aperture radius, as previously considered by \citet{Arimatsu_2011}.
The present results verify the expectation that BSCv2 point-source fluxes can be recovered from the AFASS maps by aperture photometry with proper aperture correction.

\subsubsection{Photometry with the 1-$\sigma$ Aperture without Correction}

Third, we compare our photometry results using the 1-$\sigma$ aperture without any aperture correction.
This method practically corresponds to photometry using an infinite aperture.
We employ this method of photometry, especially because photometry results must be verified not only for point sources but also for extended sources.
With this method, the AFASS-to-BSCv2 flux ratio is derived to be
$1.31 \pm	0.29$,
$1.11 \pm	0.20$,
$1.05 \pm	0.19$, and
$0.93 \pm	0.21$.
The derived ratio as a function of the BSCv2 flux for each band is also shown in Fig.\,\ref{F:corrphot}.

These measurements indicate that the BSCv2 fluxes are reproduced reasonably well by setting the photometry aperture at the ``edge" of the point sources (defined by the 1-$\sigma$ contour), while uncertainties become correspondingly higher (as the number of lower S/N pixels in the photometry aperture is larger).
However, measured fluxes in the N60 band appear to be overestimated by $31 \pm 29$\,\%.
This flux overestimate in the N60 band can be attributed to the inevitable elongation of the bright PSF core.
Because a target can be scanned in both N to S and S to N directions along the Ecliptic during multiple sky coverage,
the resulting AFASS map shows elongation of the PSF core in both N and S directions.
This effect is seen most prominently in the N60 band PSF image (Fig.\,\ref{F:psfmad}).
Hence, the surface brightness is double-counted in the elongated parts of the PSF when photometry is done by an aperture that encompasses the entire elongation.
Based on the super-PSF map in the N60 band (Fig.\,\ref{F:psfmad}), the excessively elongated parts of the PSF (e.g., $>2^{\prime}$ away from the peak) registers surface brightnesses at most a few tens of \% of the peak intensity.   
This means that at most a few tens of \% of the total flux can be accounted for in these elongated parts of the PSF (Table\,\ref{T:encircle}), corroborating the measured $\sim 30$\,\% flux overestimate in the N60 band.
In genuinely extended sources, the distribution of surface brightness is supposedly not as centrally concentrated as in point sources. 
Hence, the flux overestimates seen above as a result of the source elongation would be expected to be less severe in genuinely extended sources.
Therefore, we consider that both of these photometry methods work reasonably well.

\subsubsection{Summary: Flux Calibration in the AFASS Maps}

As demonstrated above, we verify that 
roughly $40\,\%$ flux deficiencies previously reported for point sources in the AFASS maps are caused by the adopted 90$\arcsec$-radius aperture, which typically misses about 40\,\% of the total flux distributed beyond the aperture.
We also confirm that the total flux of extended sources can be recovered without any special correction by the use of an appropriate aperture that encompasses the entire extent of the extended source.

These findings suggest that the surface brightness of the AFASS maps is calibrated correctly for non-diffuse sources as well as diffuse sources.
Therefore, with the AFASSv1 images, we hereby confirm that  
(1) photometry for point sources can be done with the aperture correction method using the correction factors presented here (Table\,\ref{T:encircle}), and
(2) photometry for extended sources can be done by summing up surface brightnesses within an appropriate aperture of the target source and converting the pixel value sum into flux by multiplying by $5.29 \times 10^{-3}\,\mathrm{([Jy]/[MJy\,sr}^{-1}]) = 2.35 \times 10^{-5}\,\mathrm{([Jy/sq.\,arcsec]/[MJy\,sr}^{-1}])\,\times 15 \times 15$\,(sq.\,arcsec; at the default pixel scale of AFASSv1).

\section{Conclusions}
\label{S:summary}

We present an independent photometric analysis of point sources detected 
in the {\sl AKARI\/} far-IR all-sky survey maps (AFASSv1; \cite{Doi_2015}).
By comparing with their corresponding flux entries in the 
the {\sl AKARI\/} bright-source catalogue ver.\,2 (BSCv2; \cite{Yamamura_2016}),
we establish that the surface brightness of the AFASSv1 maps has been calibrated properly 
for point sources and extended sources alike, 
in addition to large-scale diffuse emission against which the AFASS maps were originally absolutely calibrated.

Especially, we find that the suspected flux underestimates for point sources 
in the AFASS maps reported earlier by \citet{Arimatsu_2014} and \citet{Takita_2015}
can be mitigated by applying a proper aperture correction.
In doing so, we construct the empirical super-PSF templates ourselves (Fig.\,\ref{F:psfmad})
and derive our own point-source aperture correction factors  (Table\,\ref{T:encircle}).

The BSCv2 point-source fluxes are reproduced reasonably well 
even when direct photometry is done with a 1-$\sigma$ photometry aperture
(i.e., all the surface brightness pixel values within the aperture are summed up
to compute the source flux).
This means that direct photometry for any source 
(point, extended, and diffuse) detected in the AFASS maps 
yields a correct flux measurement when a sufficiently large aperture is used.

On the whole, the AFASS maps are useful tools to directly measure far-IR fluxes.
This is true especially for sources that are not exactly point sources (i.e., extended), because the BSC catalogue flux entries may not be appropriate for such extended sources.

\begin{ack}
This research is based on observations with AKARI, 
a JAXA project with the participation of ESA.
TU acknowledges support and encouragement by the {\sl AKARI\/} team
at the Institute of Space and Aeronautical Science
of JAXA, especially from Dr.\ Issei Yamamura.
\end{ack}


\end{document}